\def\C        {{$^{13}$C \/}}
\def\N        {{$^{14}$N \/}}
\def\NN       {{$^{15}$N \/}}
\def\NV        {{$^{14}$NV \/}}

\newcommand{\mr}[1]{\mathrm{#1}}
\newcommand{\unit}[1]{\,\mathrm{#1}}

\newcommand{\us}{\,\mu{\rm s}}

\newcommand{\rtHz}{\sqrt{\mr{Hz}}}
\newcommand{\tc}{\tau_c}

\newcommand{\Tr}{T_{1\rho}}
\newcommand{\wo}{\omega_0}
\newcommand{\wone}{\omega_1}
\newcommand{\wmw}{\omega}
\newcommand{\weff}{\omega_\mr{eff}}
\newcommand{\wrf}{\omega_\mr{rf}}

\newcommand{\Dwo}{\Delta\wo}

\newcommand{\swo}{\sigma_{\wo}}
\newcommand{\swone}{\sigma_{\wone}}

\newcommand{\Bmwone}{B_1^\mr{mw}}

\newcommand{\Brfone}{B_1^\mr{rf}}
\newcommand{\Bmin}{B_\mr{min}}
\newcommand{\uT}{\mu\mr{T}}

\newcommand{\Wone}{\Omega_1}
\newcommand{\Wrf}{\Omega}

\newcommand{\ma}{|0\rangle}
\newcommand{\mb}{|1\rangle}

\newcommand{\mx}{|x+\rangle}
\newcommand{\mix}{|x-\rangle}

\documentclass[prl,aps,twocolumn]{revtex4}

\usepackage{graphicx}
\usepackage{bm}
\usepackage{amsmath}
\usepackage{amssymb}
\usepackage{verbatim}
\usepackage[colorlinks=true, pdfstartview=FitV, linkcolor=blue, citecolor=blue, urlcolor=blue]{hyperref} 

\begin{document}

\global\emergencystretch = .1\hsize 

\title{Radio-frequency magnetometry using a single electron spin}

\author{M. Loretz, T. Rosskopf, and C. L. Degen}
  \email{degenc@ethz.ch} 
  \affiliation{
   Department of Physics, ETH Zurich, Schafmattstrasse 16, 8093 Zurich, Switzerland.
   }
\date{\today}

\begin{abstract}
We experimentally demonstrate a simple and robust protocol for the detection of weak radio-frequency magnetic fields using a single electron spin in diamond.  Our method relies on spin locking, where the Rabi frequency of the spin is adjusted to match the MHz signal frequency.  In a proof-of-principle experiment we detect a $7.5\unit{MHz}$ magnetic probe field of $\sim 40\unit{nT}$ amplitude with $<10\unit{kHz}$ spectral resolution.  Rotating-frame magnetometry may provide a direct and sensitive route to high-resolution spectroscopy of nanoscale nuclear spin signals.
\end{abstract}

\pacs{76.70.-r, 76.30.Mi, 03.65.Ta, 07.55.Ge}


\maketitle

Quantum systems have been recognized as extraordinarily sensitive detectors for weak magnetic and electric fields.  Spin states in atomic vapors \cite{budker07} or flux states in superconducting quantum interference devices \cite{ryhanen89}, for example, offer among the best sensitivities in magnetic field detection.  Trapped ions \cite{maiwald09} or semiconductor quantum dots \cite{vamivakas11} are investigated as ultrasensitive detectors for local electric fields.  The tiny volume of single quantum systems, often at the level of atoms, furthermore offers interesting opportunities for ultrasensitive microscopies with nanometer spatial resolution \cite{degen08,gierling11,grinolds12}.

Sensitive detection of weak external fields by a quantum two-level system is most commonly achieved by phase detection:
In Ramsey interferometry, a quantum system is prepared in a superposition $\frac{1}{\sqrt{2}}\left(\ma+\mb\right)$ of states $\ma$ and $\mb$, and then left to freely evolve during time $\tau$.  During evolution, state $\mb$ will gain a phase advance $\Delta\phi = \tau\Delta E/\hbar$ over $\ma$ (where $\Delta E$ is the energy separation between $\ma$ and $\mb$) that is manifest as a coherent oscillation between $\frac{1}{\sqrt{2}}\left(\ma\pm\mb\right)$ states.  These oscillations can be detected either directly or by back-projection onto $\ma$ and $\mb$.  For spin systems, which are considered here, the energy splitting sensitively depends on magnetic field $B$ through the Zeeman effect $\Delta E = \hbar\gamma B$ (where $\gamma$ is the gyromagnetic ratio), allowing very small changes in $B$ to be measured for spins with long coherence times $\tau$.

In its most basic variety, phase detection measures DC or low frequency ($\sim\unit{kHz}$) AC fields that fluctuate slower than $\tau$; in other words, the free evolution process effectively acts as a low-pass filter with bandwidth $\propto \tau^{-1}$.  Spin echo and multi-pulse decoupling sequences have been introduced to shift the detection window to higher frequencies while maintaining the narrow filter profile \cite{kotler11,delange11}.  Going to higher frequencies is advantageous for two reasons: Firstly, coherence times generally increase, allowing for longer evolution times and better sensitivities.  Additionally, spectral selectivity can be drastically improved.  While multi-pulse decoupling sequences work well for capturing $\lessapprox 1\unit{MHz}$ signals \cite{delange11}, extending this range to the tens or hundreds of MHz -- an attractive frequency range for nuclear spin detection -- is impractical due to the many fast pulses required for spin manipulation.  Moreover, the response function of multi-pulse sequences has multiple spectral windows that complicate interpretation of complex signals.

\begin{figure}[b!]
      \centering
      \includegraphics[width=0.50\textwidth]{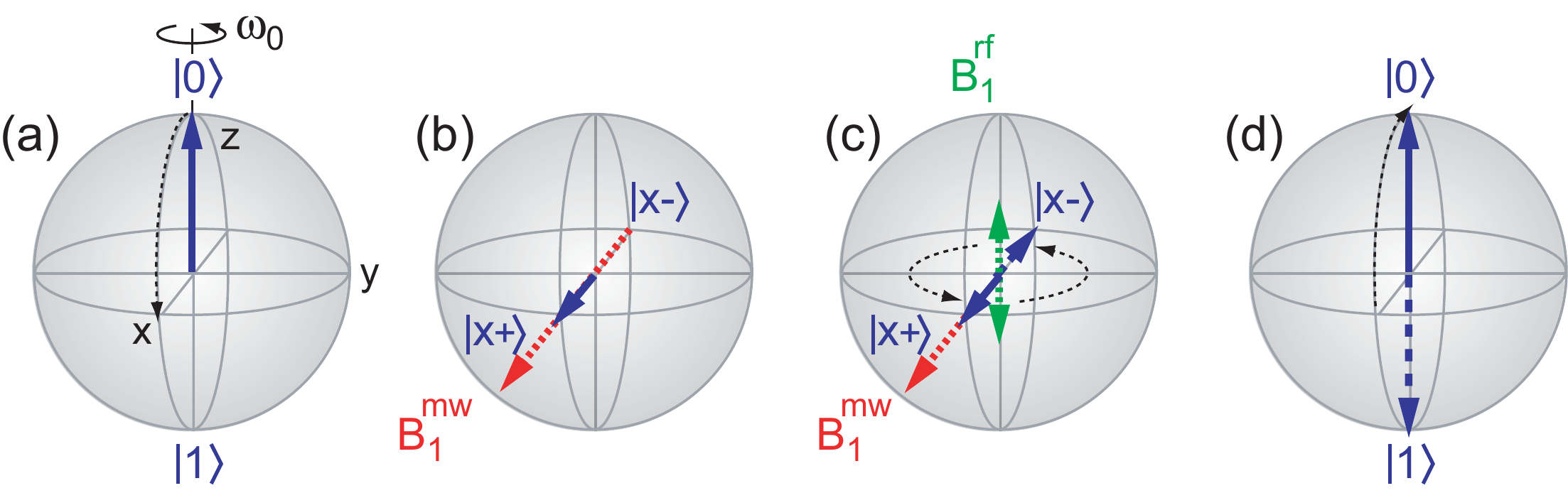}
      \caption{Illustration of rotating frame magnetometry using the Bloch sphere:  (a) An electron spin is initialized into the $\ma$ state and transferred into the coherent superposition $\mx=\frac{1}{\sqrt{2}}(\ma+e^{i\wo t}\mb)$. (b) A microwave field $\Bmwone e^{i\wo t}$ is applied in-phase with the spin precessing around the $z$ axis. (c) During sensing period $\tau$, transitions are induced between $\mx$ and $\mix=\frac{1}{\sqrt{2}}(\ma-e^{i\wo t}\mb)$ states by weak radio-frequency fields $B^\mr{rf} e^{i\Wrf t}$ if the rf frequency $\Wrf$ matches the Rabi frequency $\wone = \gamma \Bmwone$.  (d) $\mx$ ($\mix$) states are transferred to detectable $\ma$ ($\mb$) polarization.  The transition probability $\mx\leftrightarrow\mix$ is equal to the probability of measuring $\mb$.
      }
      \label{fig:bloch}
\end{figure}
Presented here is a simple and robust method to directly detect $\gg1\unit{MHz}$ magnetic signals with high sensitivity and spectral selectivity.  Our approach relies on spin locking \cite{slichter,cai11,hirose12} and is illustrated in Fig. \ref{fig:bloch}: In a spin-lock experiment, a resonant microwave field is applied in-phase with the coherent Larmor precession of the spin.  In the picture of a reference frame rotating at the Larmor frequency $\wo=\Delta E/\hbar$ (rotating frame), the microwave field appears as a constant field parallel to the spin's orientation.  In this frame of reference, the spin is quantized along the microwave field axis with an energy separation of $\hbar\wone = \hbar\gamma \Bmwone$ between states parallel and anti-parallel to the microwave field, where $\Bmwone$ is the amplitude ($\wone$ the Rabi frequency) of the microwave field.  If now an additional, weak rf magnetic field whose frequency $\Wrf$ matches the Rabi frequency is present, transitions between parallel and anti-parallel states are induced at a rate set by the magnitude $\Brfone$ of the rf field \cite{redfield55}.  Since Rabi frequencies can be precisely tuned over a wide MHz frequency range by adjusting microwave power \cite{fuchs09}, the single electron spin can act as a wide range, narrow band, and sensitive detector for rf magnetic fields.

\begin{table}[t!]
\centering
\begin{tabular}{lll}
\hline\hline
Symbol & Range & Description \\
\hline
$\wo$ & GHz & Spin Larmor frequency \\
\hline
$\wmw$ & GHz & Microwave frequency \\
$\Bmwone$ & $\mu$T-mT & Microwave amplitude \\
$\wone=\gamma\Bmwone$ & MHz & Microwave amplitude (in units\\
 &  & of frequency); Spin Rabi frequency \\
\hline
$\Wrf$ & MHz & RF frequency \\
$\Brfone$ & nT-$\mu$T & RF amplitude \\
$\Wone=\gamma\Brfone$ & kHz & RF amplitude (in units of frequency) \\
\hline\hline
\end{tabular}
\caption{List of symbols.}
\label{table:symbols}
\end{table}
In the following we consider in general terms the transition probability $p$ between $\mx$ and $\mix$ states (see Fig. \ref{fig:bloch}) in response to a coherent and to a stochastic rf magnetic probe field.  This situation is equivalent to the classical problem of a two-level system interacting with a radiation field \cite{rabi37,wang45,cummings62}.  For the case of a coherent driving field $\Brfone e^{i\Wrf t}$ oriented along the $z$-axis (see Fig. \ref{fig:bloch}), the transition probability $p$ is given by \cite{cummings62,rabi37},
\begin{equation}
p = p_0 \frac{\Wone^2}{\Wone^2+(\Wrf-\wone)^2} \sin^2\left( \sqrt{\Wone^2+(\Wrf-\wone)^2}\frac{\tau}{2} \right),
\label{eq:coherent}
\end{equation}
where $p_0\leq1$ is the maximum achievable transition probability, $\Wone=\gamma\Brfone$ is the amplitude of the rf field, and other symbols are collected in Table \ref{table:symbols}.  (If a detuning $\wmw-\wo$ were present, $\wone$ would need to be replaced by the effective Rabi frequency $\weff = \sqrt{\wone^2+(\wmw-\wo)^2}$.)  We notice that $\Brfone$ will drive coherent oscillations between states, and that the spectral region that will respond to $\Brfone$ is confined to either $\wone\pm\Wone$ or $\wone\pm 5.57/\tau$, whichever is larger \cite{supplementary}.  This corresponds to a detector bandwidth set by either power or interrogation time.

Alternatively, for stochastic magnetic signals, the transition probability can be analyzed in terms of the magnetic noise spectral density $S_{B}(\omega)$ \cite{cummings62,wang45},
\begin{equation}
p = \frac{p_0}{2} \left( 1 - e^{-\tau/\Tr} \right)
\label{eq:stochastic}
\end{equation}
where $\Tr$ is the rotating frame relaxation time \cite{slichter},
\begin{equation}
T_{1\rho}^{-1} = \frac{1}{4}\gamma^2\left[ S_{B_z}(\wone) + S_{B_y}(\wo)\right],
\label{eq:tr}
\end{equation}
 and $S_{B_z}(\wone)$ and $S_{B_y}(\wo)$ are the magnetic noise spectral densities evaluated at the Rabi and Larmor frequencies, respectively, and $z$ and $y$ are given according to Fig. \ref{fig:bloch}.  Thus, measurements of $\Tr$ for different $\wone$ can be used to map out the spectral density \cite{bylander11,bargill12}.  Eq. (\ref{eq:stochastic}) only applies for uncorrelated magnetic noise (correlation time $\tc<\tau$).  A more general expression that extends Eq. (\ref{eq:tr}) to an arbitrary spectral density is discussed in Ref. \cite{cummings62}.

Eqs. (\ref{eq:coherent}-\ref{eq:tr}) describe the general response of an ideal two-level system in the absence of relaxation and inhomogeneous line broadening.  Regarding relaxation we note that the contrast $p_0$ will be reduced to $p_0 e^{-\tau/\Tr}$ for long evolution times $\tau\gtrsim\Tr$, where $\Tr$ is the rotating frame relaxation time due to magnetic fluctuations in the sensors' environment  [Eqs. (\ref{eq:stochastic},\ref{eq:tr})].  Thus, relaxation imposes a limit on the maximum useful $\tau$, which in turn limits both sensitivity (see below) and minimum achievable detection bandwidth.

Line broadening of the electron spin resonance (ESR) transition can be accounted for by a modified transition probability $\tilde{p}$ that is averaged over the ESR spectrum $q(\wo)$ \cite{cummings62},
\begin{equation}
\tilde{p} = \int_0^\infty d\wo q(\wo) p(\wo),
\label{eq:linewidth}
\end{equation}
where $q(\wo)$ is normalized to unity.  $p(\wo)$ is given by Eq. (\ref{eq:coherent}) and depends on $\wo$ through the effective Rabi frequency $\weff = \sqrt{\wone^2+(\wmw-\wo)^2}$.  As an example, if $q(\wo)$ is a Gaussian spectrum with a linewidth sigma $\swo$ and center frequency detuned by $\Dwo=\wmw-\wo$ from the microwave frequency $\omega$, the associated linewidth of the rotating-frame spectrum is
\begin{equation}
\swone \approx \swo \left[\frac{\Dwo^2}{\wone^2}+\frac{\swo^2}{4\wone^2}\right]^{1/2} .
\label{eq:inhomogeneous}
\end{equation}
Inhomogeneous broadening of the ESR spectrum therefore leads to an associated inhomogeneous broadening of the rotating-frame spectrum that is scaled by $\frac{\swo}{2\wone}$ or $\frac{\Dwo}{\wone}$, respectively.  Since $\wone\gg\swo,\Dwo$, narrow linewidths can be expected even in the presence of a significant ESR linewidth.

Finally, we can estimate the sensitivity towards detection of small magnetic fields.  For small field amplitudes $\Wone<\tau^{-1}$ the transition probability [Eq. (\ref{eq:coherent})] reduces to $p \approx p_0 \frac{1}{4}\Wone^2\tau^2$.  Assuming that the transition probability is measured with an uncertainty of $\sigma_p$ (due to detector noise), we obtain a signal-to-noise ratio (SNR) of $\mr{SNR} = \frac{p}{\sigma_p} = \frac{p_0}{4\sigma_p}\Wone^2\tau^2$.  The corresponding minimum detectable field $\Bmin = \Wone/\gamma$ (for unit SNR) is
\begin{equation}
\Bmin = \frac{2}{\gamma\tau}\sqrt{\frac{\sigma_p}{p_0}} .
\label{eq:sensitivity}
\end{equation}
Eq. (\ref{eq:sensitivity}) outlines the general strategy for maximizing sensitivity: $\tau$ should be made as long as possible, $\sigma_p$ should be reduced (by optimizing read-out efficiency), and $p_0$ should be made as large as possible (by keeping $\tau<\Tr$ and avoiding inhomogeneous broadening) \cite{supplementary}.


%
\begin{figure}[b!]
      \centering
      \includegraphics[width=0.40\textwidth]{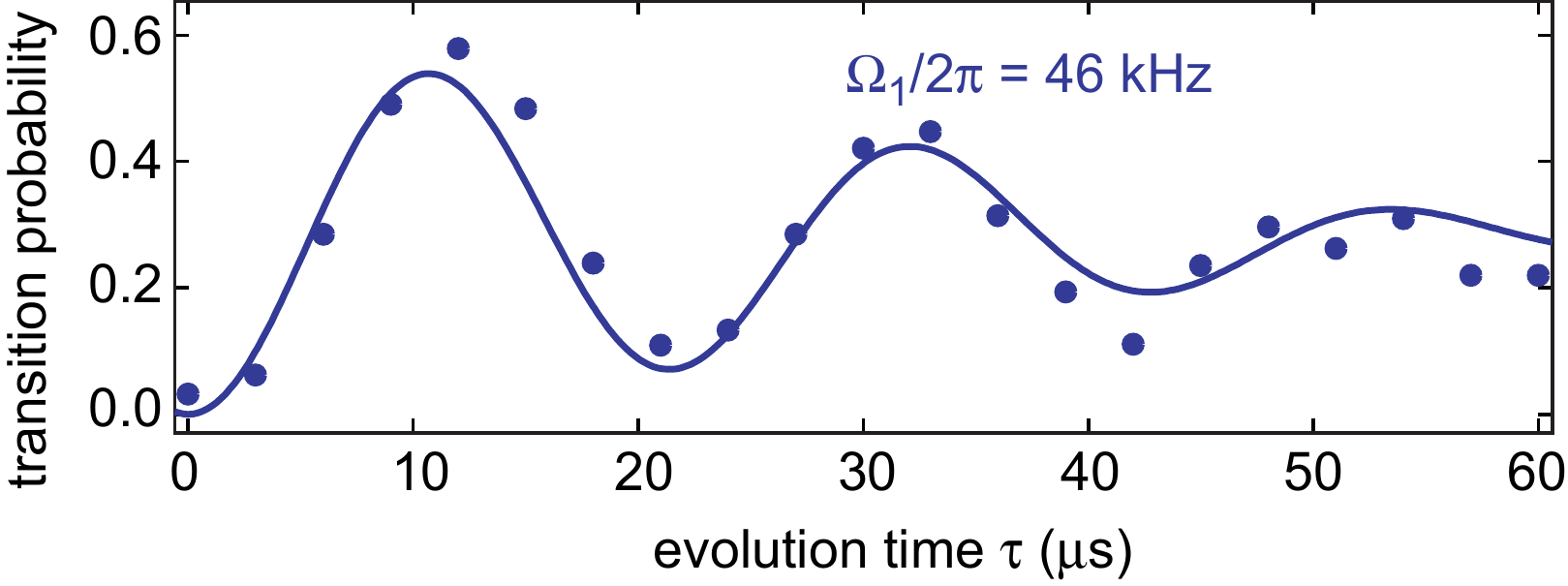}
      \caption{Coherent oscillation between $\mx$ and $\mix$ states induced by a rf probe field of $\Brfone=1.6\unit{\mu T}$.  Solid line is a fit to a decaying sinusoid \cite{supplementary}.  $\Omega$ and $\wone$ are both $2\pi\cdot7.5\unit{MHz}$.  Symbols are explained in Table \ref{table:symbols}.
      }
      \label{fig:oscillation}
\end{figure}
We demonstrate rotating-frame magnetometry by detecting weak (nT-$\uT$) rf magnetic fields using a single nitrogen-vacancy defect (NV center) in an electronic-grade single crystal of diamond \cite{oforiokai12}.  The NV center is a prototype single spin system that can be optically initialized and read-out at room temperature \cite{jelezko06} and that has successfully been implemented in high-resolution magnetometry devices \cite{balasubramanian08,rondin12,grinolds12}.  Following Fig. \ref{fig:bloch}, we initialize the NV spin ($S=1$) into the $\ma$ ($m_S=0$) state by optical pumping with a $\sim1\us$ green laser pulse, and transfer it into spin coherence $\mx$ (where $\mb$ corresponds to $m_S=+1$) using an adiabatic half-passage microwave pulse \cite{slichter}.  The spin is then held under spin-lock during $\tau$ by a microwave field of adjustable amplitude.  After time $\tau$, the state is transferred back to $\ma$ (or $\mb$) polarization, and read out by a second laser pulse using spin-dependent luminescence \cite{jelezko06}.  The final level of fluorescence (minus an offset) is then directly proportional to the probability $p$ of a transition having occurred between $\mx$ and $\mix$.  Precise details on experimental setup and microwave pulse protocol are given as Supplemental Material \cite{supplementary}.

\begin{figure}[t!]
      \centering
      \includegraphics[width=0.50\textwidth]{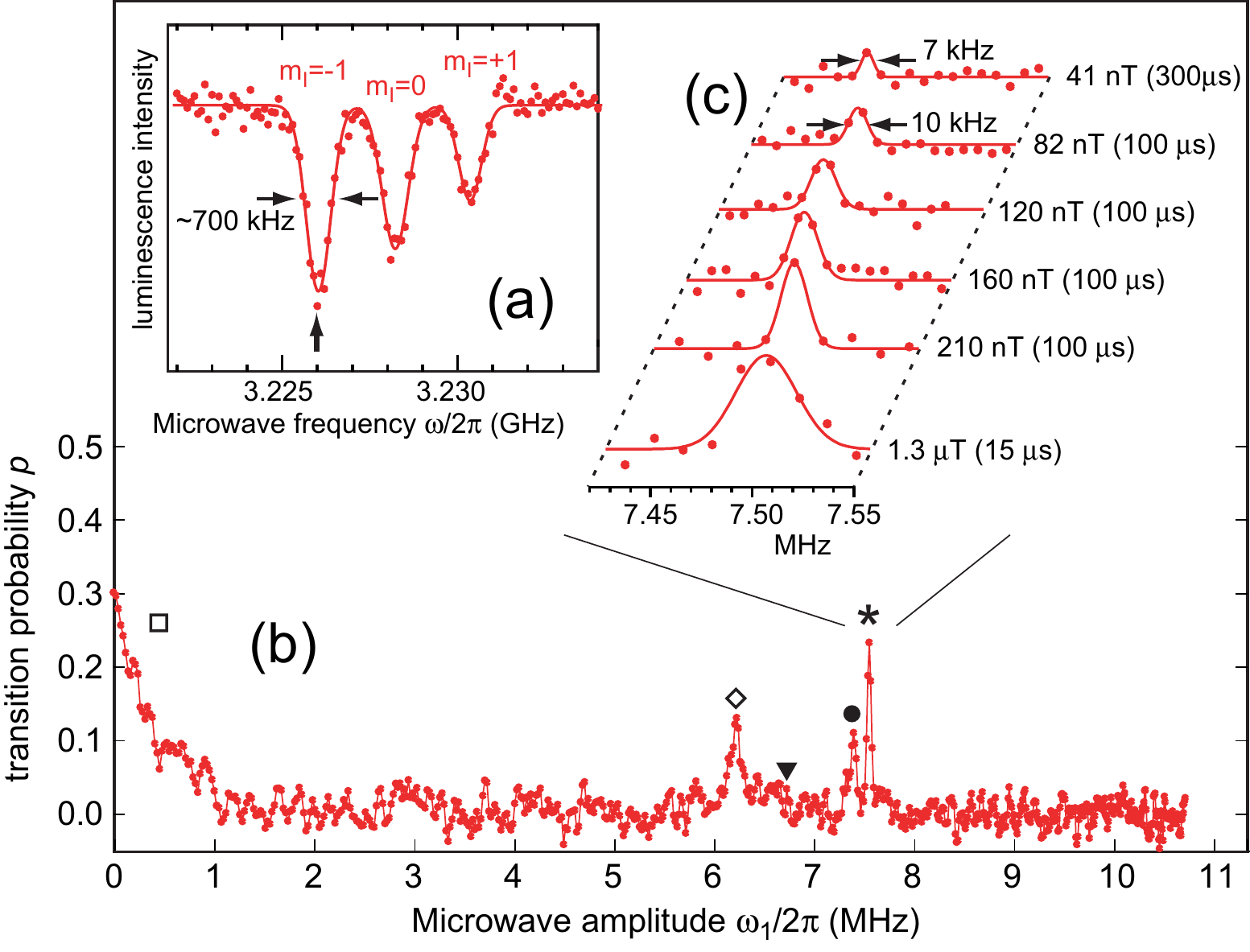}
      \caption{(a) Optically-detected electron spin resonance (ESR) spectrum of one of the NV centers used for experiments.  Dots are data, solid lines are Gaussian fits. $m_I=\pm1$, $m_I=0$ indicate hyperfine lines ($a=2.2\unit{MHz}$) associated with the \NV nuclear spin ($I=1$).  Microwave frequency used in spin-lock experiments was always centered on the $m_I=-1$ line, indicated by an arrow.  DC bias field was $17\unit{mT}$.  
      (b) Rotating-frame spectrum with 7.5 MHz probe field present (feature $\star$).  Other features are explained with Fig. \ref{fig:selective}.  Evolution time was $\tau=15\unit{\us}$.  
      (c) High-resolution spectra of the main peak ($\star$) for longer evolution times and weaker probe fields.  Dots are data, solid lines are Gaussian fits.  Linewidths are full width at half maximum.  Numbers indicate $\Brfone$ and $\tau$.  Baseline noise of the 41-nT spectrum corresponds to 8 nT ($\sigma_p=0.020$), and an integration time of 840 s per point was used.
      }
      \label{fig:rabispectrum}
\end{figure}
In a first experiment, shown in Fig. \ref{fig:oscillation}, we demonstrate the driving of coherent oscillations between parallel and anti-parallel states.  For this purpose, the microwave amplitude was adjusted to produce a Rabi frequency of $7.5\unit{MHz}$, and a small rf probe field of the same frequency was superimposed.  The transition probability $p$ was then plotted for a series of interrogation times $\tau$.  The period of oscillations allows for a precise calibration of the rf magnetic field, which in this case was $\Brfone = \Wone/\gamma = 1.65\unit{\uT}$.  While one would expect $p$ to oscillate between 0 and 1, this probability is reduced because we mainly excite one out of the three hyperfine lines of the NV center (see below).  The decay of oscillations is due to inhomogeneous broadening of the ESR linewidth and a slight offset $\Omega-\wrf$ between RF and Rabi frequencies \cite{supplementary}.

Fig. \ref{fig:rabispectrum} presents a spectrum of the transition probability up to microwave amplitudes $\wone/2\pi$ of $11\unit{MHz}$ with the same rf probe field present.  A sharp peak in transition probability is seen at 7.5 MHz (marked by $\star$), demonstrating that the electron spin indeed acts as a spectrally very selective rf magnetic field detector.  Inset (c) plots the same 7.5 MHz peak for longer evolution times and weaker probe fields, revealing that for long $\tau$, fields as small as about $40\unit{nT}$ ($1.1\unit{kHz}$) can be detected and linewidths less than $10\unit{kHz}$ (0.13\%) are achieved.
For comparison, the detection bandwidth [$2\times 5.57/\tau$, see Eq. (\ref{eq:coherent})] is $3.2\unit{kHz}$ for the 41-nT-spectrum and $8.8\unit{kHz}$ for the 82-nT-spectrum, in reasonable agreement with the experiment.  The $\sim700\unit{kHz}$ linewidth of the ESR transition [Fig. \ref{fig:rabispectrum}(a)] translates into an inhomogeneous broadening of about $18\unit{kHz}$ [Eq. (\ref{eq:inhomogeneous})], which is somewhat higher than the experiment (since the ESR linewidth is likely overestimated).  The narrow spectra together with little drift in line position furthermore underline that power stability in microwave generation (a potential concern with spin-locking) is not an issue here.

\begin{figure}[b!]
      \centering
      \includegraphics[width=0.50\textwidth]{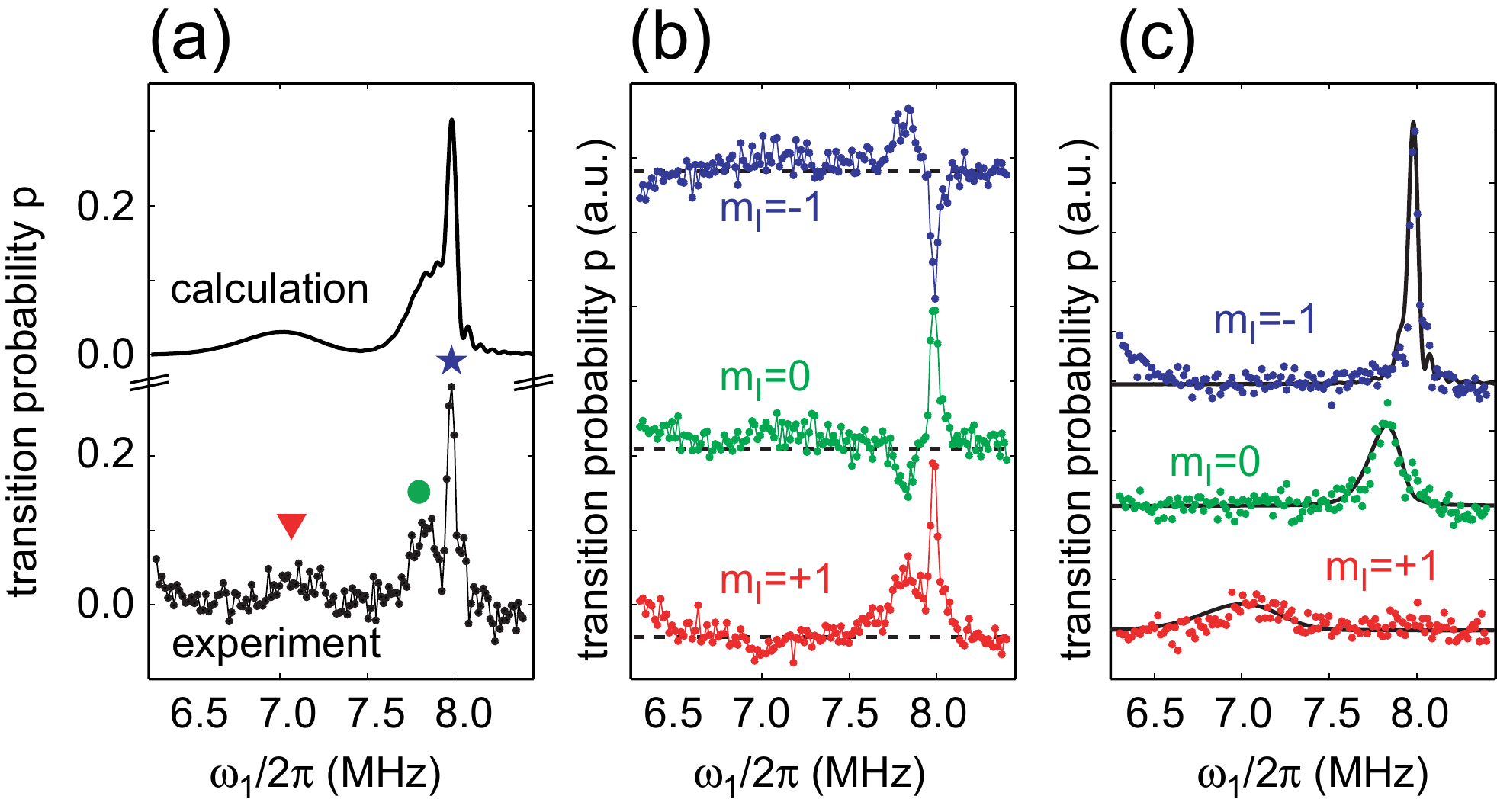}
      \caption{(a) Rotating-frame spectrum showing the three peaks associated with $m_I=-1 (\star)$, $m_I=0 (\bullet)$, and $m_I=1 (\blacktriangledown)$.  Solid line is a calculation based on Eq. (\ref{eq:linewidth}) and the EPR spectrum shown in Fig. \ref{fig:rabispectrum}(a). Parameters are identical to Fig. \ref{fig:rabispectrum} except for $\Wrf/2\pi=8\unit{MHz}$.
      (b) Spectra as-recorded using spin state selection.
      (c) Linear recombination of pure-state spectra from data in (b). Solid line is the calculated response.
      }
      \label{fig:selective}
\end{figure}
Several additional features can be seen in the spectrum of Fig. \ref{fig:rabispectrum}.  The increase in $p$ below $1\unit{MHz}$ (marked by $\square$) can be attributed to nearby \C diamond lattice nuclear spins ($I=\frac{1}{2}$, 1\% natural abundance) with hyperfine couplings in the 100's of kHz range, causing both spectral broadening and low frequency noise.  The feature ($\diamond$) appearing at $\sim 6\unit{MHz}$ is of unknown origin; since it is absent for NV centers composed of the \NN nuclear isotope it is probably related to the nuclear quadrupole interaction of \N \cite{supplementary}.  The peak at $\sim 7.3\unit{MHz}$ ($\bullet$) finally is a replica of the main $7.5\unit{MHz}$ peak ($\star$) associated with the $m_I=0$ nuclear \NV spin sublevel:
Since the microwave field excites all three hyperfine lines [see Fig. \ref{fig:rabispectrum}(a)], the rotating-frame spectrum is the stochastic thermal mixture of three different Larmor transitions with different effective Rabi frequencies.  Only two out of three peaks are visible in Fig. \ref{fig:rabispectrum}; all three peaks can be seen in a higher resolution spectrum shown in Fig. \ref{fig:selective}(a).  This presence of hyperfine lines is undesired, as it can lead to spectral overlap and generally complicates interpretation of the spectrum.

In Fig. \ref{fig:selective} we show how this complexity can be removed using spin state selection \cite{duma03}.  For this purpose, we invert the electronic spin conditional on the \N nuclear spin state before proceeding with the spin-lock sequence.  Conditional inversion is achieved by a selective adiabatic passage over one hyperfine line.  In the spectrum this leads to selective inversion of peaks associated with that particular nuclear spin state.  Fig. \ref{fig:selective}(b) shows the resulting spectra for all three sublevels.  By linear combination of the three spectra (or by subtraction from the non-selective spectrum) we can then reconstruct separate, pure-state spectra for each $m_I$ sublevel.  Fig. \ref{fig:selective}(c) shows that spin state selection is very effective in removing the hyperfine structure in the spectrum.  We note that other schemes could also be used, such as initialization of the nuclear spin by optical pumping \cite{jacques09} or more general spin bath narrowing strategies \cite{cappellaro12}.

\begin{figure}[t!]
      \centering
      \includegraphics[width=0.35\textwidth]{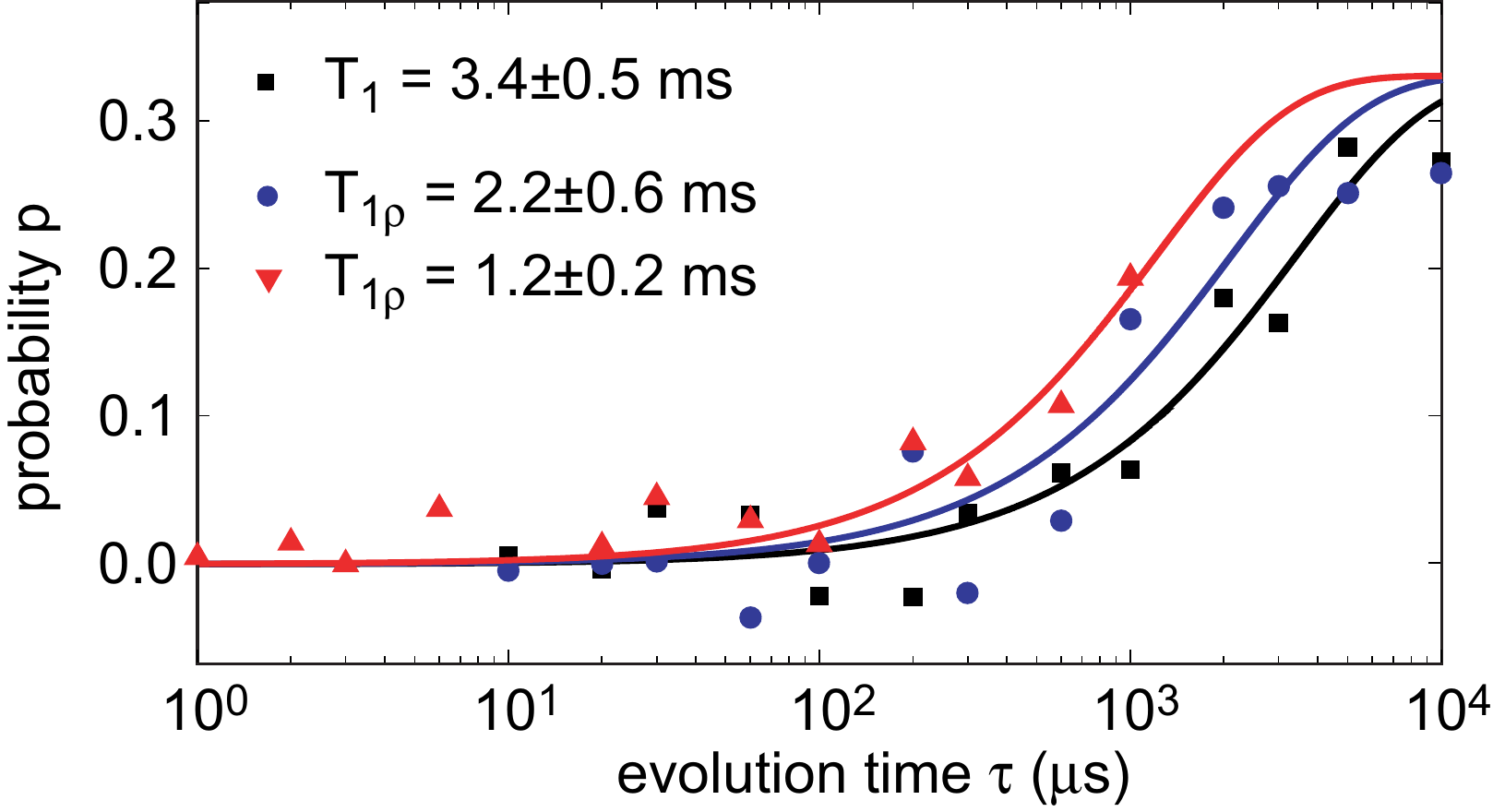}
      \caption{Relaxation time measurements for bulk (black squares and blue circles) and shallow implanted (5 nm deep, red triangles) NV centers show approximately $T_1$-limited behavior.
      }
      \label{fig:tr}
\end{figure}
Finally, we have determined the baseline magnetic noise spectral density $S_B(\omega)$ for two representative NV centers using relaxation time measurements.  Fig. \ref{fig:tr} plots $T_1$ and $\Tr$ decay curves for a bulk and a shallow-implanted ($\sim5\unit{nm}$) NV center \cite{oforiokai12}.  From the $T_1$ measurement we infer 
$S_B^{1/2}(\wo) = \sqrt{2/(\gamma^2 T_1)} \approx 0.14\unit{nT/\rtHz}$ (per magnetic field orientation), evaluated at $\wo/2\pi=3.2\unit{GHz}$.  From $\Tr$ measurements we obtain $S_B^{1/2}(\wone) \approx 0.20-0.30\unit{nT/\rtHz}$, evaluated at $\wone/2\pi=7\unit{MHz}$.  Thus, for these experiments we conclude that $S_B^{1/2}(\wone)$ is similar, if slightly higher than $S_B^{1/2}(\wo)$, and measurements are approximately $T_1$-limited.  Since $T_1$ itself is likely limited by thermal phonons and could be enhanced to $<1\unit{pT/\rtHz}$ by going to cryogenic temperatures \cite{jarmola12}, there is scope for further improvement at lower temperatures.

In conclusion, we have demonstrated how a single electronic spin can be harnessed for radio-frequency magnetic field detection with high sensitivity and excellent spectral resolution.  Our protocol relies on spin-locking and is found to be robust and simple, requiring a minimum of three microwave pulses.  Although the radio-frequency range addressed in our demonstration experiment was limited to roughly $0-11\unit{MHz}$ by efficiency of microwave delivery, it is easily extended to several hundred MHz using more sophisticated circuitry, such as on-chip microstrips \cite{fuchs09}.

We anticipate that rotating-frame magnetometry will be particularly useful for the detection and spectral analysis of high-frequency signals in nanostructures, such as in small ensembles of nuclear and electronic spins.  For example, the magnetic stray field of a single proton spin at 5 nm distance is on the order of 20 nT \cite{degen08}. 
These specifications are within reach of the presented method and engineered shallow diamond defects \cite{oforiokai12,ohno12}, suggesting that single nuclear spin detection could be feasible.  In contrast to other nanoscale magnetic resonance detection methods, such as magnetic resonance force microscopy \cite{poggio10} single electron spin sensors are ideally suited for high-resolution spectroscopy applications because they operate without a magnetic field gradient.

The authors gratefully acknowledge financial support through the NCCR QSIT, a competence center funded by the Swiss NSF, and through SNF grant $200021\_137520 / 1$.
We thank K. Chang, J. Cremer, R. Schirhagl, and T. Schoch for the help in instrument setup, sample preparation and complementary numerical simulations.  C. L. D. acknowledges support through the DARPA QuASAR program.

\end{document}



\begin{center}
{\Large Supplemental Material to the Manuscript\\ ``Radio frequency magnetometry using a single electron spin''}
\vspace{0.5cm} \\
M. Loretz$^{1}$, T. Rosskopf$^{1}$, C. L. Degen$^{1}$ \\
\end{center}

{\small\noindent
$^1$Department of Physics, ETH Zurich, Schafmattstrasse 16, 8093 Zurich, Switzerland. \\
}


\section{Experimental setup}

\subsection{Optical and microwave components}
Experiments were carried out at room temperature on a home-built confocal microscope.  Single NV centers were excited by a $<3\unit{mW}$, 532 nm laser pulse gated by an acousto-optic modulator in a double-pass arrangement.  Luminescent photons were collected by an avalanche photo diode over an effective filter bandwidth of 630-800 nm, and counted using a standard PCI counter card (National Instruments, NI6602).

Microwave pulses were generated by an arbitrary waveform generator (Tektronix AWG5002C, 600 MS/s, 14 bits) at an IF of 100 MHz and upconverted to the desired $\sim 3.2\unit{GHz}$ using an I/Q mixer (Marki Microwaves, IQ-1545) and a low-phase-noise LO (Phasematrix, Quicksyn FSW-0020).  The microwave power level was adjusted during numerical synthesis of the arbitrary waveform.  The microwave signal was amplified by a linear power amplifier (Minicircuits ZHL-16W-43+) and the MHz rf probe field (HP 33120A) added using a bias-T (Customwave CMCH0674A).  Microwaves were delivered by passing current through a lithographically patterned stripline in close proximity ($<100\unit{\um}$) to the sample and terminated in a high-power $50\unit{\Omega}$ terminator.  From the narrow spectra observed ($<$10 kHz at 8 MHz Rabi frequency) we conclude that microwave power was stable to better than 0.5\% over an entire experiment ($\sim$hours).  Rabi frequencies were independently calibrated using a series of Rabi nutations.  The absolute calibration error was on the order of $\pm 3\%$.  For better absolute accuracy, an rf probe field of known frequency can be used as reference signal. 

\subsection{Pulse-timing diagram}
\begin{figure}[b!]
      \centering
      \includegraphics[width=0.80\textwidth]{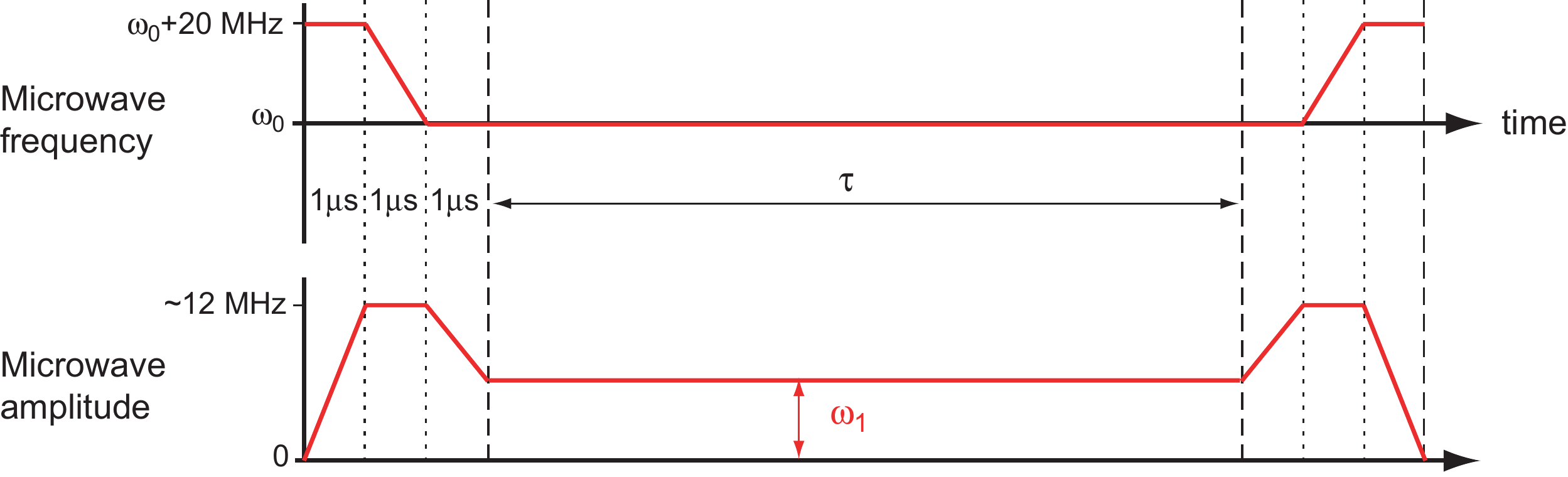}
      \caption{Pulse-timing diagram for spin-lock experiments}
      \label{fig:pulsescheme}
\end{figure}
Fig. \ref{fig:pulsescheme} depicts the pulse-timing diagram used for spin lock experiments.  At the start of each detection sequence, the spin is initialized into the $m_s=0$ state using a $1\unit{\us}$ green laser pulse (not shown).  A sequence of three linear frequency or amplitude sweeps of $\sim1\unit{\us}$ duration each is used to adiabatically transfer the spin into $\mx$ state.  The reverse sequence is used to transfer $\mx$ ($\mix$) states back to $\ma$ ($\mb$) polarization.  The polarization is detected using a second $\sim1\unit{\us}$ green laser pulse (not shown).  Along with each signal, two reference signals were recorded to calibrate the luminescence of the $m_s=0$ and $m_s=+1$ states.  The overall efficiency of the adiabatic sweep sequence was $>95\%$.  The entire sequence was repeated millions of times until typically $10'000$ photons were collected per point.

\section{Transition probability}
\label{sec:transitionprobability}

The transition probability may be analyzed for situtations where the excitation is coherent (infinite coherence time $\tc$), strongly correlated ($\tc\gg\tau$), or uncorrelated ($\tc\ll\tau$), where $\tau$ is the evolution time (spin lock duration).

\subsection{Coherent excitation ($\tc=\infty$)}

The transition probability of a two-level system with energy spacing $\wone$ subject to a coherent, harmonic driving field of frequency $\Wrf$ and amplitude $\Wone$ is given by \cite{cummings62}
\begin{eqnarray}
p & = & p_0 \frac{\Wone^2}{\Weff^2} \sin^2(\Weff\tau/2) \label{eq:coherent} \\
  & \approx & \frac{p_0}{4} \Wone^2\tau^2 \quad\quad(\mr{for\ weak\ fields,}\ \Wone\tau\ll\pi/2) \label{eq:coherent}
\end{eqnarray}
where $p_0\leq 1$ is the maximum possible transition probability and $\Weff = \sqrt{\Wone^2 + (\Wrf-\weff)^2}$.  This is Eq. (1) in the main manuscript.

\subsection{Stochastic excitation, correlated ($\tc\gg\tau$)}

This is the situation of a driving field with a slowly varying amplitude.  Since $\tc\gg\tau$, we can assume that the amplitude $\Wone$ is constant during interrogation time $\tau$ but varies between interrogation periods.  If we assume that $\Wone$ has a normal distribution (for example, it is generated by a large number of independent fluctuators, such as precessing nuclear spins), and neglecting detuning ($\Weff=\Wone$), the transition probability is
\begin{eqnarray}
p & = & \int_{-\infty}^{\infty} d\Wone p(\Wone) \frac{1}{\sqrt{2\pi}\Wrms} e^{-\frac{\Wone^2}{2\Wrms^2}} \\
  & = & \int_{-\infty}^{\infty} d\Wone p_0 \sin(\Wone\tau/2)^2 \frac{1}{\sqrt{2\pi}\Wrms} e^{-\frac{\Wone^2}{2\Wrms^2}} \\
  & = & \frac{p_0}{2} \left(1-e^{-\frac{1}{2}\Wrms^2\tau^2}\right) \\
  & \approx & \frac{p_0}{4}\Wrms^2\tau^2 \quad\quad(\mr{for\ weak\ fields,}\ \Wrms\tau\ll\pi/2)
\label{eq:correlated}
\end{eqnarray}
$\Wrms$ is the rms amplitude of the fluctuating magnetic field.  A nominal rotating-frame relaxation time $\Tr$ is found by setting $e^{-\frac{1}{2}\Wrms^2\Tr^2} = 1/e$, thus $\frac{1}{2}\Wrms^2\Tr^2 = 1$ and $\Tr=\sqrt{2}/\Wrms$.

\subsection{Stochastic excitation, uncorrelated ($\tc\ll\tau$)}

This is the situation derived in classical magnetic resonance textbooks \cite{slichter}.  The transition probability is given by
\begin{equation}
p = \frac{p_0}{2}\left(1 - e^{-\tau/\Tr}\right),\ \mr{where}\ \Tr^{-1} = \frac{1}{4}\gamma^2S_B(\omega) .
\label{eq:tr}
\end{equation}
$\gamma$ is the electronic gyromagnetic ratio.  If the noise is centered around zero frequency and $\omega<\tc^{-1}$, as for many lifetime-limited processes, the spectral density is frequency independent, $S = 2\Wrms^2 \tc$, and the transition probability is given by 
\begin{eqnarray}
p & = & \frac{p_0}{2} \left( 1 - e^{-\frac{1}{2}\Wrms^2\tc\tau} \right) \\
  & \approx & \frac{p_0}{4}\Wrms^2\tc\tau \quad\quad(\mr{for\ weak\ fields,}\ \Wrms\tau\ll\pi/2)
\end{eqnarray}
%

\section{Linewidth}

Linewidth in rotating-frame magnetometry is determined by three parameters: Spin lock duration $\tau$, magnetic field power in the vicinity of the detection frequency, and inhomogeneous broadening of the electron spin resonance transition.  Each parameter imposes a limit on the linewidth, and the largest of the three contributions will set the experimentally observed linewidth.

\subsection{Time-limited linewidth}

Short $\tau$ lead to a wider detection bandwidth (in analogy to lifetime-limited processes).  If we denote detector bandwidth by $b$, and define it as twice the frequency offset $\Omega - \wone$ for which the probability function $p$ [Eq. (\ref{eq:coherent})] is reduced to 50\% of the maximum value,
\begin{equation}
\frac{p(\Omega-\wone=b/2)}{p(\Omega-\wone=0)} = 0.5
\end{equation}
we can numerically solve for $b$ to find:
\begin{equation}
b = 5.5680\, \tau^{-1}
\end{equation}
This relation applies for weak magnetic fields ($\Wone<\tau^{-1}$) where power broadening is small.  An example for an evolution-time-limited linewidth is shown by the blue curve in Fig. \ref{fig:coherentresponse}, where $\Wone=\frac{\pi}{4}\tau^{-1}$.

\subsection{Power-limited linewidth}

\begin{figure}[b!]
      \centering
      \includegraphics[width=0.85\textwidth]{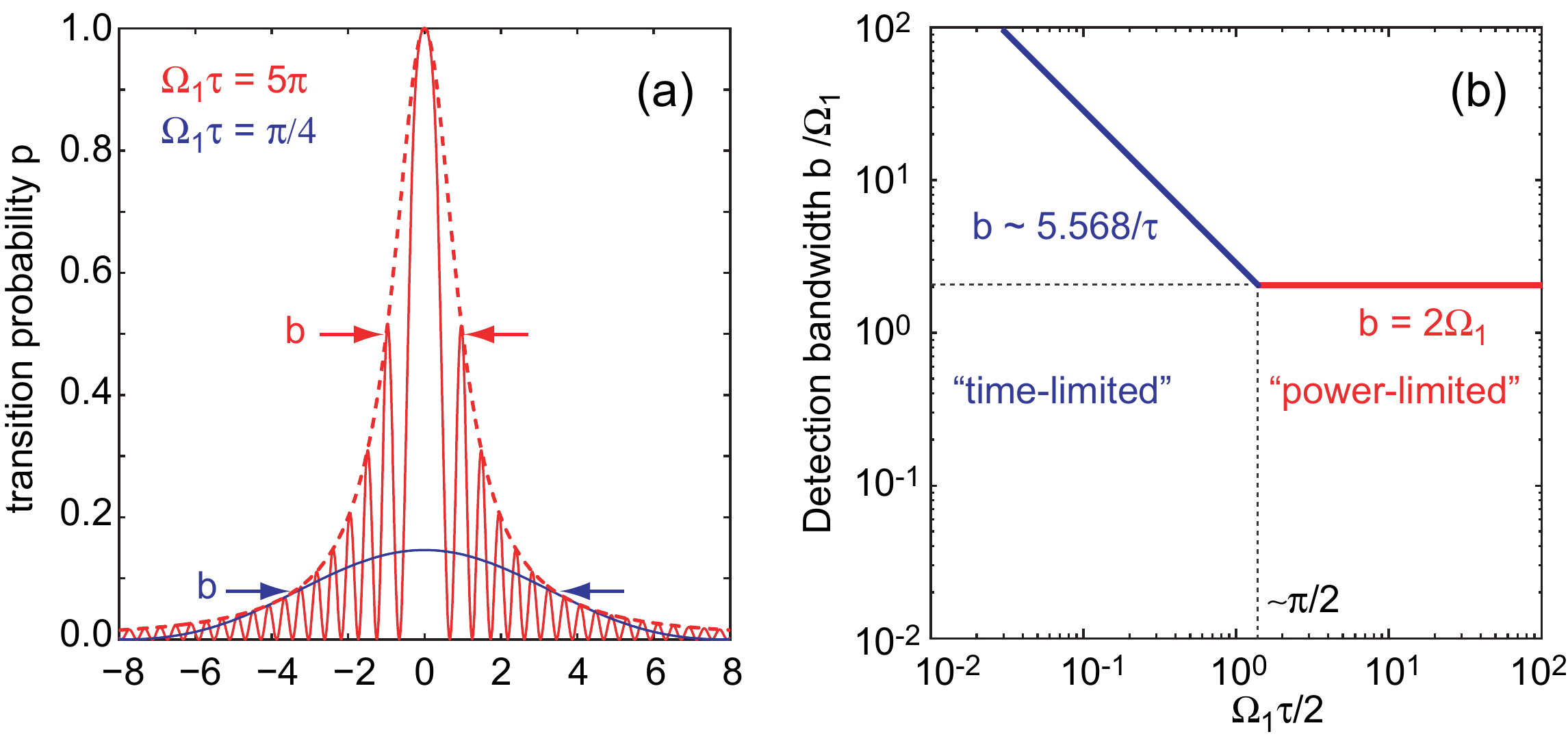}
      \caption{(a) Eq. (\ref{eq:coherent}) plotted as a function of detuning $\Omega-\weff$ for an evolution-time-limited situation (blue, $\Wone\tau=\pi/4$) and a power-broadened situation (red, $\Wone\tau = 5\pi$). Dashed red line is the envelope.  (b) Detection bandwidth $b$ plotted as a function of $\Wone\tau$, indicating the two linewidth regimes.
      }
      \label{fig:coherentresponse}
\end{figure}

Large magnetic probe fields will saturate the detector and lead to power broadening.  The condition for power broadening is $\Wone>\tau^{-1}$, and in this case, the detection bandwidth is dominated by the Lorenzian factor in $p$ [Eq. (\ref{eq:coherent})].  The power-broadened detector bandwidth is:
\begin{equation}
b = 2\Wone
\end{equation}
An example for a power-broadened linewidth is shown by the red curve in Fig. \ref{fig:coherentresponse}, where $\Wone=(5\pi)\tau^{-1}$.  The detector bandwidth is given by the full-width-at-half-maximum of the envelope (dashed curve).

\subsection{Inhomogeneously broadened linewidth}

A third line broadening mechanism is through fluctuations in the effective Rabi frequency $\weff = \sqrt{\wone^2+(\omega-\wo)^2}$, which may be caused by fluctuations in microwave power, microwave frequency, or the Larmor frequency of the electron spin.

\subsubsection{Fluctuations in microwave power}

Fluctuations in microwave power $P$ lead to fluctuations in the microwave amplitude $\wone\propto\sqrt{P}$.  Specifically, a variation of $\delta P$ in microwave power $P$ will lead to variation of $\delta\wone \approx \wone (\delta P/2P)$ in detector frequency.  Since power fluctuations can occur with microwave generation, they are a potential concern.  We have not calibrated power stability for our microwave generation, but from the fact that narrow spectra ($<10\unit{kHz}$) were observed at a detector frequency of 7.5 MHz we can deduce that power stability was better than $\delta P/P \lesssim 2\cdot(10\rm{kHz}/7.5\rm{MHz}) \approx 0.3\%$ for these experiments.

\subsubsection{Jitter in microwave frequency}

Jitter in microwave frequency is negligible.

\subsubsection{Inhomogeneous broadening of ESR linewidth}
\label{sec:linewidth}

Fluctuations in the electron spin Larmor frequency $\wo$, due to inhomogeneous broadening of the ESR linewidth, lead to an associated line broadening in the rotating-frame spectrum.  Typically, as is shown in the following, the rotating-frame line broadening is reduced by $\propto \frac{\swo}{\wone}$ compared to the original ESR linewidth, where $\swo$ is the ESR linewidth and $\wone$ the Rabi frequency.

We show this for the specific situation where the EPR spectrum $q(\wo)$ is described by a Gaussian:
\begin{equation}
q(\wo) = \frac{1}{\swo\sqrt{2\pi}} \exp\left[ -\frac{(\wmw-\wo-\Dwo)^2}{2\swo^2} \right] .
\end{equation}
%

Here, $\Dwo$ is the mean detuning and $\swo$ the sigma of the resonance. (For the \NV center, three Gaussians were added to reproduce the three hyperfine resonances).
Since $q(\wo)$ is a Gaussian, the modified transition probability $\tilde{p}$ is also approximately a Gaussian, with a peak shift $\Dwone$ given by
\begin{equation}
\Dwone = \wone - \weff \approx \wone - \wone \left(1 + \frac{\Dwo^2}{2\wone^2}\right) \approx -\frac{\Dwo^2}{2\wone} ,
\end{equation}
where $\weff = \sqrt{\wone^2+(\wmw-\wo)^2}$, and a sigma $\swone$ given by
\begin{eqnarray}
\swone^2 & = & \left(\frac{\partial\Dwone}{\partial\Dwo}\right)^2 \swo^2 + \left(\frac{\swo^2}{2\wone}\right)^2  \\
         & = & \left[\frac{\Dwo^2}{\wone^2}+\frac{\swo^2}{4\wone^2}\right] \swo^2 .
\end{eqnarray}
The above equations assume that $\Dwo,\swo\ll\wone$ (which applies to the experiments presented here).  The according inhomogeneous linewidth $\binh$ is
\begin{equation}
\binh = \sqrt{8 \ln(2)} \swone \approx 2.35 \swone
\label{eq:inhomogeneous}
\end{equation}

For example, for an ESR linewidth with $\swo/(2\pi)\approx 350\unit{kHz}$ and zero detuning ($\Dwo=0$), as well as a Rabi frequency of $\wone/(2\pi)=7.5\unit{MHz}$, the associated linewidth of the rotating frame has a sigma of approximately $\swone/(2\pi) = 7.7\unit{kHz}$ (see Table \ref{table:selective}).  The corresponding full-width-at-half-maximum value is $\binh\approx2.35\cdot7.7\unit{kHz}\approx18\unit{kHz}$.  Since we were able to observe linewidths $<10\unit{kHz}$, the original EPR linewidth is probably overestimated.


\section{Sensitivity}

Sensitivity of rotating-frame magnetometry is influenced by several parameters:  Spin lock duration $\tau$,  magnetic background noise (described by a finite relaxation time $\Tr$), efficiency of the optical read-out (including shot noise and optical contrast), and inhomogeneous broadening of the ESR transition.

\subsection{Interrogation time}

Spin-lock duration directly determines the transition probability between the two quantum states (see Section \ref{sec:transitionprobability}).  For weak fields, the transition probability is approximately [see Eq. (\ref{eq:coherent})]:
\begin{equation}
p = \frac{1}{4}p_0\Wone^2\tau^2
\end{equation}
where $p_0$ is peak contrast, $\Wone$ the amplitude of the magnetic field to detect, and $\tau$ spin lock duration.

\subsection{Magnetic background noise}

Magnetic noise at the detector is typically dominated by the fluctuating bath of spins surrounding the electronic spin sensor.  For NV centers in diamond these spins may be \C lattice nuclei, N donors, or surface impurities.  Magnetic noise at the detector is the most fundamental noise source and sets a lower limit to the minimum detectable field.

In the experiment, magnetic background noise is observed as rotating-frame relaxation, reducing $p_0$ for evolution times $\tau$ approaching $\Tr$ (according to Eq. \ref{eq:tr}). If rotating-frame relaxation needs to be taken into account, $p_0$ must hence be substituted by $p_0\rightarrow p_0 e^{-\frac{\tau}{\Tr}}$.

\subsection{Optical read-out efficiency}

The dominant measurement noise in our measurement comes from shot noise in the optical read-out of the spin state, due to the finite number of photons collected per point.  For a total number of photons collected $C$, the shot noise is $\sigma_C = \sqrt{C}$.  The total number of photons detected $C$ is
\begin{equation}
C = N r = (T/t_\mr{meas}) r
\end{equation}
where $N$ is the number of measurements, $r$ is the number of photons collected per measurement (collection efficiency), $T$ is the total measurement time, and $t_\mr{meas}$ is the duration of one measurement ($t_\mr{meas}\gtrsim\tau$).

The shot-noise limited signal-to-noise ratio is given by
\begin{equation}
\mr{SNR} = \frac{\Delta C}{\sqrt{C}} = p\epsilon\sqrt{C} \approx p\epsilon\sqrt{\frac{Tr}{\tau}}
\end{equation}
where $C$ is the number of photons detected, $\Delta C = p \epsilon C$ is the change in photon counts due to the magnetic probe field ($\Delta C \ll C$), and $\epsilon$ the optical contrast between $\ma$ and $\mb$ spin states.

\subsection{Inhomogeneous line broadening}

Inhomogeneous broadening of the rotating-frame linewidth (see Section \ref{sec:linewidth}) reduces the peak transition probability $p$ if it is the largest contribution to the rotating-frame linewidth.  Specifically, if the sensor has a detector bandwidth of $b$ (either time- or power-limited) an inhomogeneously broadened linewidth $\binh>b$, then peak probability is reduced by a factor $x=b/\binh$.  (This can be understood as a convolution between the two spectral functions with fixed areas under each curve.)

\subsection{Overall Sensitivity}

Collecting equations from Sections 4.1-4.4 and using $\Wone=\gamma\Brfone$, where $\gamma$ is the electron gyromagnetic ratio, the overall SNR for magnetic field sensing becomes:
\begin{equation}
\mr{SNR} = \frac{1}{4} x p_0 e^{-\frac{\tau}{\Tr}} \epsilon(\gamma\Brfone)^2\tau^{1.5}(Tr)^{0.5} .
\end{equation}
The corresponding minimum detectable field (for unit SNR) is:
\begin{equation}
\Bmin = \left[ \frac{1}{4} x p_0 e^{-\frac{\tau}{\Tr}} \epsilon\gamma^2\tau^{1.5}(Tr)^{0.5} \right]^{-0.5} .
\label{eq:bminshot}
\end{equation}
Alternatively, one can experimentally infer the minimum detectable field from the standard deviation $\sigma_p$ of the baseline noise of the transition probability.
\begin{equation}
\sigma_p = \frac{p_0}{4}\gamma^2\Bmin^2\tau^2 .
\end{equation}
Solving for $\Bmin$,
\begin{equation}
\Bmin = \frac{2}{\gamma\tau}\sqrt{\frac{\sigma_p}{p_0}}
\label{eq:bminexpt}
\end{equation}
For the last two equations, as well as sensitivity values reported in Table \ref{table:sensitivity}, it was assumed that $x=1$ and $\tau\ll\Tr$.

\section{Calculations and fits to experimental data}

\subsection{Coherent oscillations between $\mx$ and $\mix$ (Figure 2)}

The oscillation in Fig. 2 was fit to
\begin{equation}
p = \frac{p_0}{2} \left( 1 - e^{-t^2/(2T)} \cos(\Wone t) \right) .
\end{equation}
(This equation is equivalent to Eq. (1) in the main manuscript, except for that a Gaussian decay was added).  Here, $p_0$ is contrast, $\Wone$ is the oscillation frequency, and $T$ a Gaussian decay constant.  The fit yielded $p_0=0.56$, $\Wone/(2\pi)=46.1\unit{kHz}$, and $T=28\unit{\us}$.

The decaying sinusoid can also be directly calculated from Eqs. (1), (4) and (5) in the main manuscript for an inhomogeneously broadened ESR transition.  Specifically, we can calculate $p(\tau,\Omega)$ as a function of evolution time $\tau$ and RF frequency offset $\Omega-\wone$, and then integrate over $\Omega$ assuming a Gaussian distribution of offsets $\Omega-\wone$:
\begin{equation}
\tilde{p}(\tau) = \int_{-\infty}^\infty d\Omega
  \left( \frac{1}{\swone\sqrt{2\pi}} \exp\left[-\frac{(\Omega-\wone-\Delta\Omega_0)^2}{\swone^2}\right] \right) p(\tau,\Omega) .
\end{equation}
Here, $\swone$ is the inhomogeneous broadening (sigma) of the rotating-frame spectrum and $\Delta\Omega_0$ is the mismatch between $\Omega$ and $\wone$.  We have calculated $\tilde{p}(\tau)$ for the dataset shown in Fig. 2 in the main manuscript and find that $\swone/(2\pi)\approx 15\unit{kHz}$ and $\Delta\Omega_0\approx30\unit{kHz}$ in this measurement.

\subsection{Sensitivity (Figure 3)}

The sensitivity was calculated in two ways.  First, shot-noise limited sensitivity was calculated from experimental parameters. Second, the sensitivity was directly inferred from the standard deviation of the baseline in the transition rate spectra shown in Fig. 2(b,c).  Parameters are collected in Table \ref{table:sensitivity} and refer to Sections 4.2 and 4.3.  All numbers are per point.
\begin{table}[h!]
\centering
\begin{tabular}{lll}
\hline\hline
Quantity & Main spectrum in Fig. 2(b)	& 41-nT-spectrum in Fig. 2(c) \\
\hline
Evolution time $\tau$											&	$15\unit{\us}$					&	$300\unit{\us}$ \\
Single measurement duration $t_\mr{meas}$	& $31.4\unit{\us}$				& $316.4\unit{\us}$ \\
Total time $T$														&	145 s										& 840 s \\
Total counts $C$													&	9'700										& 10'200 \\
Number of measurements $N$								& $4.6\ee{6}$							& $2.7\ee{6}$ \\
Photons per measurement	$r$								& 0.0021									& 0.0039 \\
Contrast $\epsilon$												& 0.31										& 0.31 \\
Probability $p_0$													& 0.45										& 0.45 \\
Baseline noise $\sigma_p$									& 0.029										& 0.020 \\
\hline
$\Bmin$ from shot noise	[Eq. (\ref{eq:bminshot})]					& 170 nT									& 10 nT \\
$\Bmin$ from baseline noise	[Eq. (\ref{eq:bminexpt})]			& 200 nT 									& 8 nT \\
\hline\hline
\end{tabular}
\caption{Experimental parameters and magnetic field sensitivity $\Bmin$ for spectra in Fig. 3(b,c).  }
\label{table:sensitivity}
\end{table}
%

\subsection{Linewidth (Figure 3)}

Experimental linewidths indicated for the 41-nT and 82-nT spectra in Fig. 3(c) are determined using a Gaussian fit and are reported as full width at half height (FWHH = $2.3548\sigma$).  Corresponding calculated linewidths reported in the text are determined by numerically finding the FWHH as explained in Section 3.1.  The EPR-linewidth induced line broadening (Section 3.2.2) for these spectra is on the order of 8 kHz, see Table \ref{table:selective}.

\subsection{Spin-state-selection spectra (Figure 4)}

The transition probability shown in Figures 4(a,c) was calculated as function of $\wone$ for each hyperfine line separately, and the three probabilities then added (this is valid as long as there is no strong overlap between features).

The transition probability associated with a single hyperfine line was calculated according to Section 3.2.2.  The mean detuning $\Dwo$ and linewidth $\swo$ were obtained from Gaussian fits to the ODMR spectrum shown in Fig. 3(a).  These values along with detuning $\Dwone$ and linewidth $\swone$ of the rotating-frame spectrum are collected in Table \ref{table:selective}. 
\begin{table}[h!]
\centering
\begin{tabular}{llllll}
\hline\hline
\N sublevel & $\wone$	& $\Dwo$ & $\swo$ & $\Dwone$ & $\swone$ \\
\hline
$m_I=-1$ & 8 MHz & -0.5 MHz & 350 kHz & -16 kHz & 22 kHz \\
$m_I=0$ & 8 MHz & 1.7 MHz & 350 kHz & -180 kHz & 74 kHz \\
$m_I=+1$ & 8 MHz & 3.9 MHz & 350 kHz & -950 kHz & 170 kHz \\
\hline
--- & 8 MHz & 0 MHz & 350 kHz & 0 kHz & 7.7 kHz \\
\hline\hline
\end{tabular}
\caption{Line shift $\Dwone$ and linewidth $\swone$ for experiments shown in Fig. 2(a) and Fig. 4(c) in the main manuscript.  Lowest row is for an ideal situation of zero detuning.}
\label{table:selective}
\end{table}
%

\subsection{$T_1$ and $T_{1\rho}$ relaxation times (Figure 5)}

$\Tr$ relaxation time measurements used the pulse-timing diagram as shown in Figure \ref{fig:pulsescheme} with a variable spin lock duration $\tau$.  $T_1$ relaxation time measurements used no microwaves at all and simply consisted on an initialization of the spin by a first green laser pulse, and readout of the spin state by a second green laser point after time $\tau$.  Relaxation time measurements were fitted to
\begin{equation}
p = p_0 \left(1-e^{-t/T}\right) ,
\end{equation}
where $p_0 = 1/3$ is contrast and $T$ is the respective relaxation time ($T_1$ or $\Tr$).

\section{Origin of the 6-MHz feature}

This section collects experimental data on the peak that appeared between 5.5 and 6.2 MHz in Figs. 3 and 4.  While we do not know the exact cause and mechanism resulting in this feature.
%
%

Experimental observations:
\begin{itemize}
\item The feature appeared at frequencies between approximately 5.2 MHz and 6.3 MHz, depending on the measurement.
\item The feature was observed on three independent \NV centers.  It was not observed on any \NNV center.
\item The feature is due to a coherent source, evidenced by oscillation similar to Fig. 2.  Two measured oscillation frequencies were about 34 kHz and 17 kHz, respectively. 
\item The feature is present regardless whether the auxiliary RF signal line was connected or disconnected.
\item The feature is related to the $m_I=-1$ nuclear spin state (as evident from Fig. 4). We have not observed any peaks related to $m_I=0,+1$, but cannot exclude them with the given SNR.
\end{itemize}

Based on these observations we suspect that the feature is related to the \N nuclear spin of the NV center.  One possible mechanism is through cross-terms in the Hamiltonian that result from a slight vector misalignment of the $10-20\unit{mT}$ bias field.  Another mechanism is through the small perpendicular component of the dipolar hyperfine interaction \cite{felton09}.

\section{Additional measurements}

\begin{figure}[t!]
      \centering
      \includegraphics[width=0.70\textwidth]{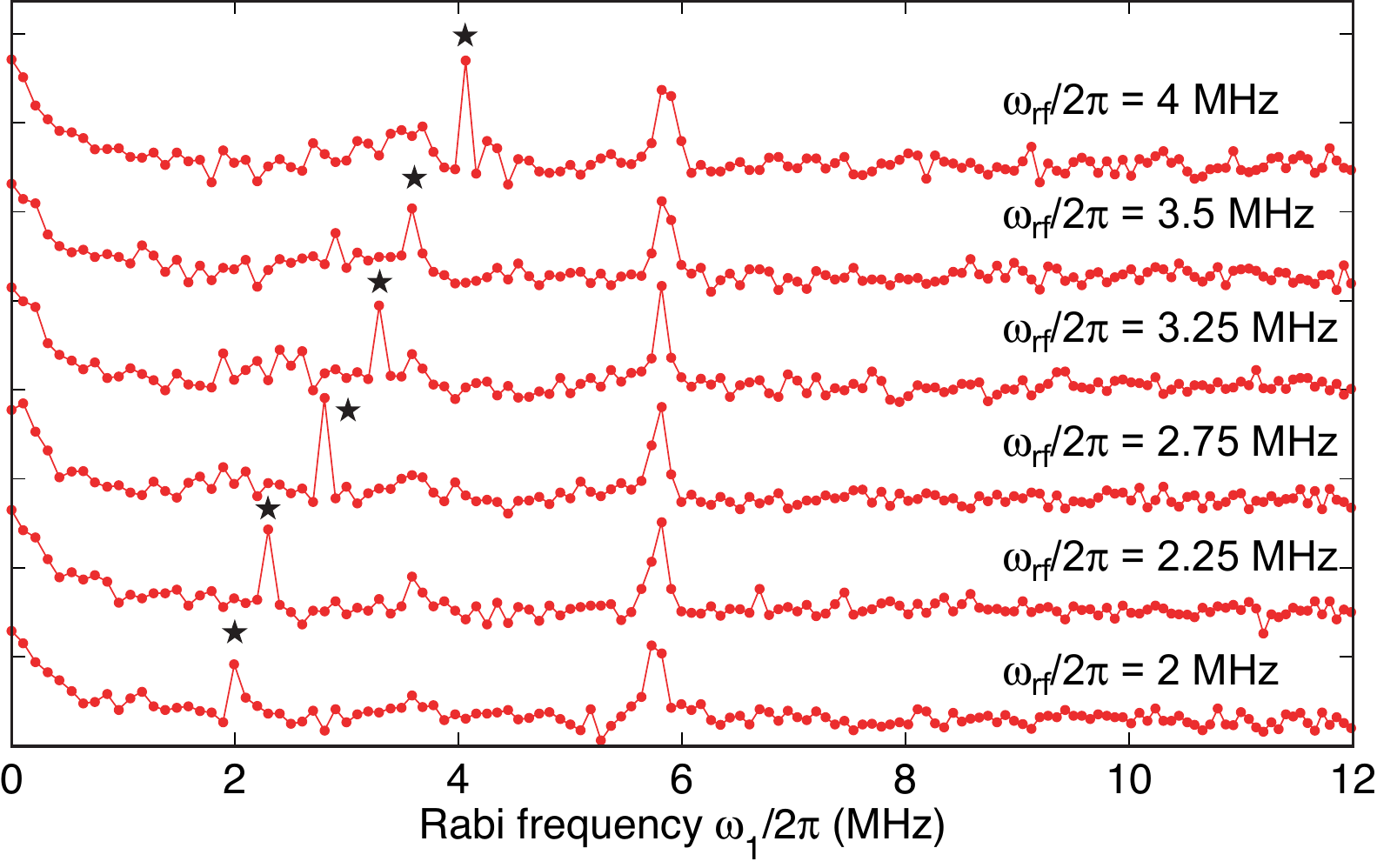}
      \caption{Series of rotating-frame spectra for rf probe field frequencies $\Wone/2\pi$ between 2 and 4 MHz.  Signal peaks are marked by $\star$.
      }
      \label{fig:waterfall}
\end{figure}
Fig. \ref{fig:waterfall} shows a series of spectra for rf probe field frequencies $\Wone/2\pi$ between 2 and 4 MHz, complementing the spectra shown in the main manuscript that were obtained at higher $\Wone/2\pi$.  Magnitude of rf field was between 1 and 2 $\mu$T.
